\documentclass[11pt]{article}
\usepackage[utf8]{inputenc}
\usepackage[T1]{fontenc}
\usepackage{mathptmx}
\usepackage[margin=1.1in]{geometry}
\usepackage{microtype}
\usepackage{hanging}
\usepackage{setspace}
\usepackage[hidelinks]{hyperref}
\usepackage{xurl}
\usepackage{textcomp}

\title{Free Speech and Artificial Intelligence\thanks{Preprint. Forthcoming in Mark Satta, Étienne Brown, and JP Messina (eds.), \emph{Philosophy and Free Speech: An Introduction}. London: Routledge, 2027.}}
\author{Étienne Brown\\ \small University of Ottawa}
\date{}

\begin{document}
\maketitle

\begin{abstract}
\noindent Philosophers and legal scholars are engaged in debates about the implications of artificial intelligence for freedom of expression. This chapter analyzes the free speech issues raised by two distinct AI technologies: social media recommendation algorithms and conversational AI (i.e., chatbots powered by large language models). The first part shows that, through their recommendation algorithms, social media platforms control the dynamics of speech visibility in the digital public sphere, making algorithmic recommendation relevant to the philosophy of free speech. The second part turns to conversational AI. It discusses both the reasons for granting or withholding speech rights to artificial agents and users' right to receive information, which may render specific forms of chatbot regulation illegitimate. Throughout, the chapter also considers whether social media platforms or AI developers hold corporate speech rights. Its general aim is to raise rather than settle questions that arise from the rapid development of AI technologies.
\end{abstract}

\onehalfspacing

\section{Introduction}

The release~of ChatGPT in November 2022 marked the beginning of the era of readily and widely available machine-generated speech. Since then, the capabilities and popularity of \textbf{large language~models} (LLMs) have grown so much that, when people speak of artificial intelligence (AI), they often primarily think of chatbots such as ChatGPT, Claude, or Gemini. Given the short history of publicly available LLMs, it might therefore come as a surprise that legal scholars have discussed the relationship between AI and freedom of expression for more than a decade (Solum 1992; Benjamin 2013; Volokh and Falk 2012; Wu 2013). Granted, many contributions on this topic focus on slightly older AI technologies, such as search engines and social media \textbf{recommendation algorithms}. Still, the last fifteen years of legal scholarship have been marked by the publication of several papers that all focus on a similar question:~``Is machine speech legally protected speech?''

While the current chapter touches on this question, it also goes beyond it. Specifically, I propose a philosophical examination of the free-speech challenges posed by two influential AI technologies: social media recommendation algorithms and conversational AI (i.e., LLM-powered chatbots). As we will see, algorithmic recommendation raises questions in direct continuity with those addressed by Howard and Fisher in the previous chapter of this volume. If users have a right to free speech on social media platforms, what is the scope of this right? Does it protect them against forms of algorithmic content moderation that fall short of removal? Do the owners of social media platforms engage in expressive conduct when they implement a suite of algorithms that amplify some pieces of content, but not others? In the first part of this chapter, my central claim will be that recommendation algorithms are relevant to the philosophy of free speech as they are key drivers of speech visibility (and invisibility) in the digital public sphere.

In the second part, I will address the question increasingly raised by contemporary legal scholars: ``Is machine speech legally protected speech?'' To answer this question, we first need to identify whose rights are potentially at stake: the rights of AI conversational agents, those of AI users, and those of AI developers. I will also distinguish between two scenarios that raise distinct political-philosophical questions: one in which a human dialogues with an LLM, and another in which a user publishes AI-generated texts, images, or videos elsewhere on the internet, for instance, on a social media platform.

\section{Free Speech and Social Media Recommendation Algorithms}

\subsection{Algorithmic Reach and the Value of Speech}

Recommendation algorithms -- also called recommender systems or recommenders -- are among ``the most visible success stories of AI in practice'' (Jannach et al. 2022). For billions of users, the algorithms that rank and distribute content on apps such as Facebook, X, Instagram, TikTok, YouTube, and Reddit have become part of everyday life\textquotesingle s infrastructure. Simply put, recommenders are suites of algorithms that determine what users see \emph{first} and \emph{next} on a digital platform (Brown 2025). They have enabled an internet experience where users can read, watch, or listen to content without having to search for it. Through scrolling or swiping, users are endlessly exposed to content without any necessary deliberate or reflective input on their part. However, in many instances, users are not passive recipients of content; they also seek to communicate with others by reacting, writing, or speaking. When they do, recommendation algorithms matter. If your goal is to be seen or heard by a large audience on a given social media platform, whether you will reach that goal essentially depends on AI systems that determine how many algorithmic impressions your speech deserves.

If we keep these remarks in mind, the implications of recommendation algorithms for freedom of expression are not hard to see. Let us begin by reflecting on the nature of our individual right to free speech. Within liberal frameworks, free speech is traditionally understood as a negative right that protects its bearers from governmental interference. As social media platforms are not governments, the application of this right to the relationship between a social media platform and its users is not straightforward. Nevertheless, many believe that such an application is justified given that mainstream platforms control a large share of the public's attention and even exercise law-like power over their users' speech (Lazar 2025). If so, then such users might have a moral right to be protected against certain forms of \textbf{private censorship}. For instance, Chris Bousquet (2025: 61) has recently suggested that ``viewpoint-based moderation by social media companies'' violates users' right to free expression by undermining ``subjects' equal opportunity for political influence.'' In the previous chapter of this volume, Howard and Fisher also note that social media platforms might have a duty to serve as free speech zones, grounded either in their voluntary public commitments or in the very fact that they have become \emph{de facto} public squares where public communication occurs. Developing this line of argument, Matthew Kramer (2021: 58-59) writes that:

\begin{quote}
\ldots social-media platforms such as Facebook, Twitter, and YouTube have become public fora. Although the companies that create and run those platforms are not morally obligated to sustain them in existence at all, the role of controlling a public forum morally obligates each such company to comply with the principle of freedom of expression while performing that role. No constraints that deviate from the kinds of neutrality required under that principle are morally legitimate.
\end{quote}

For the sake of discussion, let us then assume that large social media platforms have a duty to respect their users' right to free expression. The important question for our purposes regards the scope of this duty. When discussing violations of the right to free speech, many political philosophers focus on \emph{removal}. The scenarios they envision are ones in which (i) a user posts a piece of content on a given platform and (ii) either this piece is censored from the platform or the offending user's account is banned. Yet, platforms often engage in moderation interventions that fall short of removal (Gillespie 2022). A widely discussed phenomenon is that of \textbf{shadowbanning} (also called algorithmic demotion), which describes a situation in which user-generated content is not removed from a platform, but rather made ineligible for algorithmic recommendation. In this case, the content remains searchable, but it won't appear in the feeds of users who do not actively search for it. In other words, social media firms now have the means to add \emph{degrees of invisibility} to user-generated content without removing it from their platforms. This raises an interesting but difficult question: how many degrees of invisibility must be added to a given piece of content before we consider that it is morally equivalent to removal? If social media platforms have a moral duty to refrain from engaging in certain forms of removal, do they also have a moral duty to refrain from engaging in certain forms of shadowbanning? To answer these questions, let us consider some of the central speech-related interests that the right to freedom of expression is meant to protect. It will then be easier to see that, just like removal, making someone's speech imperceptible significantly threatens some of these interests.

Seana Shiffrin (2014: 87) has argued that speech is a necessary condition of becoming oneself as it allows one to develop a mental life that contributes ``to a distinctive perspective that embodies and represents each individual's separateness as a person.'' Beyond self-knowledge, speech also enables us to present an image of ourselves that helps others relate to us appropriately (Brown 2023). In doing so, it helps us forge meaningful relationships in which we foster human values such as intimacy, friendship, and reciprocity. To a large extent, ``It is through the medium of speech that individuals disclose themselves, appear before others, are recognized and affirmed by them, and acquire a sense of who they are'' (Parekh 2012: 43).

Speech alone is rarely sufficient to effectively pursue these interests. To meaningfully interact with others, we need to \emph{communicate} with them, which itself presupposes that our utterances are heard and understood. This is what philosophers of language have traditionally called \emph{uptake} (Austin 1975). Matteo Bonotti and Jonathan Seglow (2022: 517) forcefully illustrate this point in a recent article:

\begin{quote}
\ldots the negative right to free speech and the opportunity to exercise it are analytically distinct. But we need to consider the meaning of a free speech right from the perspective of a potential speaker who wishes to communicate something to others, or to influence them, or make a request of them, or to engage in deliberation with them, and so on {[}\ldots{]} The point of free speech is to address others and possess some traction in doing so; a person's speech, to have value for them, needs to have the potential for uptake. Without this, free speech, for some at least, is a liberty without worth.
\end{quote}

Here, a key idea is that speakers won't be able to ``address others'' unless they are heard. Metaphorically put, very few speakers are interested in expressing themselves alone in a desolate forest; we primarily wish to speak with interlocutors or in front of an audience. At the very least, the interests we satisfy by speaking alone represent only a small fraction of those we seek to defend when we claim our right to free speech. Freedom of expression derives most of its value from the fact that it promotes \emph{communicative} as opposed to \emph{merely expressive} interests.

On social media platforms, speakers pursue communicative interests. Many users try to convince others that their speech is worth hearing, hoping to influence public opinion, feel validated, build communities, or make financial gains. Now, making a speaker's speech invisible through algorithmic demotion is just as likely to frustrate these communicative interests as removing it altogether. Consequently, my suggestion is that algorithmic demotion sufficiently limits freedom to count as a form of speech interference that, like removal, needs justification. On social media platforms, the difference between removing a piece of content and making it \emph{very hard to see} is not significant enough to consider that removal is morally suspect whereas algorithmic demotion is not.

Of course, this does not mean that all social media users are owed a large audience. First, such speakers might not need access to large audiences to fulfill all their speech-related interests (Miller 2021). Second, giving them one would unduly constrain other users' freedom not to pay attention to content they find uninteresting or objectionable. As we will see below, it might also unduly constrain social media platforms' corporate right to expression. Regardless, my point is that algorithmic sanctions, such as demotion, sometimes affect speakers' ability to satisfy their interests in ways similar to censorship. Consequently, philosophers who oppose platforms' prerogative to remove content as they please have reason to believe that they should not demote content at their discretion either.

Moreover, we can imagine cases in which social media users are unable to pursue certain speech-related interests due to low algorithmic reach, even if their speech has not been moderated. A recommender system might repeatedly predict that a user's speech is unlikely to be engaging and, for this reason, limit its diffusion. In this case, arguing that the platform is wronging the user is harder, as it does not directly interfere with the user's speech. As we have seen, the right to free speech is typically considered a purely negative right that protects users from interference, not one that permits speakers to request assistance from others when their speech is barely heard.~

Nevertheless, there is one democratic requirement that might lead us to view a lack of algorithmic reach as a moral problem. In democratic societies, all citizens should have a meaningful opportunity to influence political outcomes. Before the rise of social media platforms, political philosophers already worried that this requirement was rarely fulfilled. They even suggested that, to meet it, democratic citizens need to be given their fair share of the public's attention. For instance, Ronald Dworkin (2010: viii) writes:~

\begin{quote}
It may be objected in most democracies that right now, the right to free speech has little value for many citizens, ordinary people with no access to great newspapers or television broadcasts have little chance to be heard. This is a genuine problem. It may be that genuine free speech requires more than just freedom from legal censorship {[}\ldots{]} we must try to find other ways of providing those without money or influence a real chance to make their voices heard.
\end{quote}

At first glance, social media platforms offer a new opportunity for speakers to voice their opinions: anyone with enough time and digital literacy can log in to a platform of their choice and share their thoughts with others. If platforms mark an improvement over the previously gatekeeper-heavy status quo, it might seem an ill fit to demand that the companies that provide these further opportunities take on additional speech-related duties. At the same time, observers have pointed out that, on social media, attention is heavily concentrated. As Arvind Narayanan (2023) explains, on social media platforms, ``those who already have a high reach, whether earned or not, are rewarded with more reach. For example, the top 1\% of authors on Twitter (now X) receive 80\% of tweet views.'' If this is true, social media platforms might enable people to \emph{express} themselves without giving them a real chance to be heard (Brown 2026).

Perhaps the burden of doing so should not fall on platforms. After all, what democracy requires is that all citizens be given a meaningful opportunity to influence the political process \emph{in society at large}, not on a given social media platform. Furthermore, as I discuss in the next section, platforms might also have expressive rights that limit what the public can reasonably ask of them. Nonetheless, the findings that platforms can moderate at will, algorithmically control the fate of digital speech, and sharply concentrate attention might motivate states to limit their size. Think, for instance, of antitrust measures or a public grants approach designed to establish a variety of online public forums powered by different recommendation algorithms. Without enabling states to control platforms' choice of a particular recommender system---such control would pose a risk of abuse---we can support initiatives that aim to give citizens more meaningful opportunities to be heard.

\subsection{Algorithmic Recommendation as Expressive Conduct}

When governments attempt to control the behavior of social media platforms, are they within their rights to do so? Recently, officials have argued that such efforts would violate these platforms' right to freedom of expression. In \emph{Moody v. NetChoice, LLC}, 603 U.S. 707 (2024), the U.S. Supreme Court ruled against Texas and Florida's attempts to prevent social media platforms from engaging in certain types of content moderation. In these cases, the two states argued that platforms violated viewpoint neutrality and discriminated against conservatives, but the Court rejected their plea. Central to the Court's reasoning was the idea that defining and implementing moderation guidelines constitutes expressive conduct. As Justice Kagan (26) explained:~

\begin{quote}
When the platforms use their Standards and Guidelines to decide which third-party content those feeds will display, or how the display will be ordered and organized, they are making expressive choices. And because that is true, they receive First Amendment protection.
\end{quote}

In various legal contexts, corporations have a right to free speech comparable to that of natural persons. However, in \emph{Netchoice}, the argument went further. Judges argued that when platforms moderate, they are, in a sense, speaking.~Consequently, government-mandated moderation would violate their First Amendment rights. This is not the first time the U.S. Supreme Court has intervened to protect information providers' expressive choices. In a 1974 case -- \emph{Miami Herald Publishing Co. v. Tornillo}, 418 U.S. 241 (1974) -- it ruled against Florida's attempt to coerce newspapers into printing responses from political candidates who had been criticized by the paper.

Significantly for our purposes, the Court's decision in \emph{Netchoice} also raised the question of whether implementing recommendation algorithms constitutes expressive conduct. This is hardly surprising: strictly speaking, it is not primarily content moderation, but algorithmic recommendation, that determines how users' feeds ``will be ordered and organized.'' While the Supreme Court declined to rule on whether platforms speak through their recommendation algorithms, one lower-court judge has interpreted Kagan as arguing that they do. In \emph{Anderson v. TikTok, Inc.}, 116 F.4th 180 (3d Cir. 2024), the U.S. Court of Appeals for the Third Circuit reasoned that \emph{TikTok}'s decision to design and implement its popular \emph{For You Feed} counts as speech. According to Judge Patty Shwartz (9), ``TikTok makes choices about the content recommended and promoted to specific users, and by doing so, is engaged in its own first-party speech.''~

If Shwartz is correct, then limiting a platform's choice of recommendation algorithms threatens its legal right to freedom of expression. Certainly, Shwartz could be wrong (Goldman 2024), and the very idea that corporations should have extended speech rights has long been criticized for enabling opportunism (Schauer 2002). In the U.S., corporations often invoke their First Amendment rights to pursue goals that have little to do with expression, such as selling medical patients' prescription records or refusing to display the full price of airline tickets to customers. Yet, if we accept the claim that (i) corporations have a right to free speech and (ii) these rights cover the choice of content moderation rules, the conclusion that such rights also extend to the choice of a particular recommender architecture is hard to avoid. On digital platforms, recommenders impact user experience just as much as moderation does.

One way to challenge this idea is to argue that content moderation amounts to editorial control, whereas choosing a recommender system does not. In a recent contribution, West, Novelli, Taddeo, and Floridi (2025: 52) develop such an argument:

\begin{quote}
Because moderation involves selecting content with the explicit intent of conveying values in ordinance with a platform's community guidelines, courts have found that it conveys meaning, is interpretable, and achieves these aims through a discrete set of acts, thus meeting the three criteria for pro­tection. We argue, however, that curation does not rise to the same standards of editorial judgment. Curation is not expressive but \emph{predictive}, as it operates by statistically modeling users based on their behavior and presenting con­tent to maximize some success metrics, such as prediction accuracy or user engagement.
\end{quote}

The idea in question is that predictive algorithms cannot be as expressive as non-predictive ones. Does this idea withstand philosophical scrutiny? I doubt it. Simply put, predictive goals can be expressive. Consider the following example. Although current social media algorithms attempt to predict \emph{user engagement}, scholars have argued that platforms should implement \textbf{bridging algorithms} that aim to amplify content likely to elicit positive responses from people who usually disagree with one another (Ovadya and Thorburn 2023). Advocates of bridging argue that it reduces political polarization and promotes mutual understanding and trust on digital platforms. Now, if a social media CEO implements bridging algorithms because she believes that reducing political polarization is a laudable democratic goal, this decision is arguably just as expressive as choosing to implement a particular moderation policy (for instance, one that prohibits hate speech). The fact that recommenders are predictive tools does not invalidate this claim. In fact, the key question worth asking is not ``Are you trying to predict?'' but rather ``\emph{What} are you trying to predict and \emph{why}?'' Second, even designing purely engagement-based algorithms could reasonably be seen as a form of expressive conduct. Imagine, for instance, a local newspaper editor telling his staff: ``In this newsroom, we do one thing and one thing only. We publish articles that people find engaging, regardless of content. Our readers work 9 to 5, and when they pick up their newspaper, they deserve to have a good time.'' Would this choice not be expressive? Would it be any less expressive than those of an editor who prioritizes intellectually enriching articles over merely engaging ones? My view is that it is not. At the very least, this remains an open question.

\subsection{Algorithmic Recommendation and the Moral Foundations of Free Speech}

A third reason why algorithmic recommendation matters philosophically is that it can obstruct the pursuit of goals that freedom of expression is intended to serve. If one instrumentally values free speech because it facilitates the formation of true beliefs or promotes intellectual autonomy, then one can legitimately worry that engagement-optimizing algorithms amplify content that does precisely the opposite. For instance, in his contribution to this volume, Jonathan Seglow contends that such algorithms prioritize content that is ``wild, sensationalist, and arresting over that which is more considered, informative, and challenging.'' Consequently, it ``does not involve exposing us, in Millian fashion, to the widest possible range of views and to encourage us to test and contest them.'' In a different chapter of this volume, Seb Bishop notes that social media algorithms distribute ``hate speech, polarizing speech, and misinformation,'' which he considers ``a threat to intellectual autonomy.'' Consequently, some proponents of autonomy-based justifications of free speech are likely to argue in favor of a more stringent regulation of bad speech or of the algorithms that rank it highly (Brown 2023).

Beyond truth and autonomy, the preceding argument naturally extends to philosophical conceptions according to which freedom of expression is primarily valuable because it is an essential instrument of democratic self-governance. By way of example, Grafanaki (2018: 131) notes that current recommenders amplify content that is ``very different than the one that promotes democratic culture.'' On social media platforms, there is ``no exchange of ideas in the marketplace that eventually leads to the truth, or any dialogue in a Meiklejohnian town meeting where decision-making is collective process.'' Instead, ``mutual understanding between groups becomes harder, leading to group polarization.'' Nikolas Kirby (2026) has also offered a democracy-based argument for social media regulation. In his view, engagement optimization algorithmically drives affective polarization, thus contributing to democratic erosion. Here, it is worth noting that the causal link between algorithmic recommendation and polarization is hard to establish (Liu et al. 2025). But if Kirby is right, what is at stake is the very capacity of democratic citizens to trust their rulers and political adversaries, which is a precondition of ``the very continuance of democracy, defined by compliance with its most basic norms,'' such as respecting the outcome of a fair election (24). As platforms flout their duty not to knowingly contribute to \textbf{affective polarization}, states can legitimately regulate them to ensure they do not.

Whether free speech is valued because it promotes truth, autonomy, or democracy, the structure of the argument remains the same (Messina 2020). If one instrumentally values free speech because it serves a given goal, then the empirical finding that current recommenders hinder the pursuit of this goal provides a reason to regulate them. Of course, such a reason might not be overriding. Even if algorithms hinder the pursuit of important moral goals, corporations might have a protected right to implement a particular recommender architecture. According to some, empowering the government to regulate platforms also entails an unacceptable risk of abuse (Messina 2024).

\section{Free Speech and AI chatbots}

\subsection{Do LLMs have a right to free speech?}

Imagine two possible regulatory approaches to large language models. In the first scenario, a government concerned about AI safety requires all startups of a certain size to implement government-sanctioned content moderation policies: LLM-powered chatbots should not provide users with detailed instructions for building weapons, generating dangerous conspiracy theories, or encouraging self-harm. In the second scenario, a government preoccupied with the putative liberal bias of popular chatbots orders startups to ensure that the political information they offer users reflects diverse viewpoints. Although hypothetical, these scenarios are not far-fetched. Researchers and journalists increasingly worry that, when it comes to conspiracy theories or self-harm, chatbots are not sufficiently moderated (FitzGerald 2025), and some have publicly criticized governments for leaving AI ``to the free market wild west'' (Hinsliff 2025). Conversely, some free speech scholars argue that AI chatbots are subject to overly strict content moderation. In their recent snapshot of corporate moderation policies for generative AI, Calvet-Bademunt and Mchangama (2024: 4) argue that these policies ``do not align with the benchmark international human rights standards,'' ``go significantly beyond the legitimate interests that justify speech restrictions,'' and ``may be biased regarding specific topics'' such as transgender athletes or the amount of power held by white Protestants in the U.S. Although the authors fall short of recommending governmental regulation, not all political observers will exercise the same restraint (as demonstrated, for example, by the Florida and Texas bills against viewpoint-based moderation discussed in the previous section).~

If governments coerced AI startups into implementing \emph{stricter} or \emph{more lenient} moderation guidelines, would that impede the speech rights of any stakeholder? To answer this question, it helps to consider whose speech rights may be at stake. Consider first a possibility that seems preposterous at first glance: that AI conversational agents possess speech rights. Of course, the very term ``AI conversational agent'' presupposes that we can reasonably assign agency to AI, a view that eminent philosophers have recently defended. For instance, Christian List (2021: 1219) writes that ``we can certainly think of AI systems {[}\ldots{]} as intentional agents of a non-biological sort.'' In his view, AI systems that operate relatively autonomously would qualify as intentional agents if they possessed (i) representational states, (ii) motivational states, and (iii) ``a capacity to interact with its environment on the basis of these states.'' Writing in 2020, List does not explicitly discuss LLMs and remains noncommittal about the empirical question of which AI system (if any) currently meets his three conditions. Writing from a functionalist perspective, he nonetheless suggests that ``agency admits at least three different kinds of hardware: biological, as in the case of human beings and non-human animals; electronic, as in the case of robots and other AI systems; and social, as in the case of group agents'' (1221).

Regarding rights, List expresses skepticism about the idea that an AI system could have non-derivative rights, that is, the kind of rights we grant to beings with ``intrinsic moral significance,'' for instance, human beings and other animals (1234). The key argument is that current computers lack phenomenal consciousness, which is ``a necessary (though perhaps not by itself sufficient) condition for having non-derivative rights'' (1236). Focusing their reflection on the capacity for well-being, Basl and Bowen (2020: 296) offer a similar argument:

\begin{enumerate}
\def\labelenumi{\arabic{enumi}.}
\item
  AI is a potential rights-holder if it is a bearer of well-being.
\item
  AI is a bearer of well-being only if it is conscious.
\item
  AI is not conscious.
\item
  So, AI cannot be said to have rights.
\end{enumerate}

Such an argument warrants two remarks. First, some philosophers emphasize that it remains deeply uncertain whether current AI systems are or can be conscious. As Schwitzgebel (2025: 99) notes, neither experts nor society at large know whether AI systems are conscious, and this situation is unlikely to change any time soon: ``Given the unsettled theoretical landscape and the extraordinary difficulty of assessing consciousness in strange forms of intelligence, uncertainty will be justified.'' Competing philosophical theories of consciousness will yield contradictory judgments. Second, there might be reasons to attribute rights to AI systems even if we are uncertain that they meet the conditions for intrinsic moral significance. Perhaps we should do so \emph{just in case} they do. After all, the moral costs of our mistakes could be very high. Alternatively, List explains that rights attributions can be derivative. In certain cases, we grant rights to entities not because they have intrinsic moral significance, but because doing so protects the interests of those who clearly do. Think, for instance, of corporate rights.~

Even if we grant rights to AI systems, that does not settle the question of whether we should grant them \emph{speech} rights. Several legal scholars have debated this issue. In an influential paper, Massaro and Norton (2016: 1172) contend that AI systems ``may be disconnected enough and smart enough to say that the speech they produce is theirs, not ours, with no human creator and director in sight.'' Arguing against this view, Tim Wu (2013: 1498) invokes the \textbf{functionality doctrine} to suggest that, like search engines, speaking machines are mere communication tools that ``facilitate the communications of another person.'' Crucially, they lack the capacity to identify with the information they handle and, for this reason, should not be granted a right to freedom of expression. In response, Garvey argues that Wu's line of reasoning overestimates the cognitive capacities of humans. Drawing on Jonathan Haidt's (2001) work on moral judgment, he notes that ``at least some human speech and action is the product of unconscious emotions'' rather than rational, reflective deliberation (Garvey 2022: 974). From his perspective, humans unduly downplay the similarities between human speech and machine speech. In both cases, a significant amount of expressed speech is not reflectively endorsed by the speaker. This leads him to conclude ``some types of AI are eligible for free speech protection under the First Amendment'' (953).

Here, philosophy can help us see things more clearly. One significant idea that Garvey's reflection overlooks is that humans possess rational capacities they do not always exercise when speaking. Even if people's speech is sometimes guided by their emotions, human beings still have the \emph{capacity} to reflectively endorse (or reject) their utterances before or after they speak. By way of contrast, we can reasonably doubt that AI systems have the capacity to reflectively endorse anything, so Garvey's proposed analogy fails. Insisting on the differences between AI systems and human beings, some philosophers of language also deny that the former possess the same speech powers as the latter. For example, Butlin and Viebahn (2025) have argued that AI systems cannot make assertions. In their view, only sanctionable speakers have the power of assertion, but an LLM is not ``capable of being sanctioned by agents with which it interacts'' (968). Indeed, if an LLM produces unreliable outputs, we can reduce our trust in it as we do with malfunctioning thermometers, but this does not amount to \emph{sanctioning} it. When we reduce our trust in a malfunctioning thermometer, we do not expect it to modify its behavior, as it is not sensitive to our feedback and lacks the interests that would make our negative disposition towards it meaningful. Importantly, the fact that we do not trust the thermometer is also not bad for it. According to Butlin and Viebahn, LLMs and thermometers are similar in at least one important way. Even if LLMs might be sensitive to our negative feedback, losing credibility is not bad for them. This itself suffices to deny them the capacity to \emph{assert} anything at all.

To sum up, one popular philosophical view of artificial intelligence holds that LLMs (i) are not conscious, (ii) lack the same rational and linguistic capacities as humans, and (iii) lack speech-related interests (since they lack the capacity to have interests in the first place). For this reason, we should not grant them a non-derivative right to freedom of expression. To avoid this conclusion, one could always defend an ``output-only'' position, according to which whatever generates speech speaks, and whatever speaks should be granted speech rights. My suggestion is that this position would either be unconvincing -- if the speaking machine lacks interests, what are its rights for? -- or amount to a derivative attribution of rights. And the very idea that we could grant AI systems a \emph{derivative} right to free speech invites us to consider who else's interests and rights are at stake when humans interact with LLMs.~

\subsection{AI users' right to receive information}

Beyond conversational AI agents, human users would also be affected by a state's decisions to regulate AI-powered chatbots. In human-chatbot interactions, a complicating factor is that regulations would neither prevent a human user from speaking nor make it harder for them to reach an audience (as was the case in our discussion of algorithmic demotion above). Even if a chatbot is programmed not to answer specific queries, human users remain free to ask it anything. Of course, regulations could prevent them from receiving an answer. Would this be an infringement of their rights?

One relevant idea is that the people's right to freedom of expression includes the right to seek and receive information, which some AI regulations could violate. For instance, Marciel (2023) contends that democratic citizens possess a right to information, which imposes obligations on both journalists and governments. In his view, such a right is grounded in two distinct political interests: (i) a citizen's interest ``in becoming well informed, in order to be able to competently advance her own political views'' and (ii) her interest ``in her fellows being well informed, so that they do not advance harmful or unjust policies'' (366-367). Furthermore, he underscores that citizens' right to information is essential to preserve democracy, as an uninformed citizenry can easily fall prey to demagogues and ``end up favoring policies that undermine basic democratic values'' (367).

While the right to information entails a correlative duty for democratic states ``to promote the conditions under which good journalism can flourish,'' Marciel recognizes that its practical implications are hard to pin down (377). One option is to interpret it as grounding negative duties for governments, which should refrain from regulating information providers too strongly. This \emph{laissez-faire} view is defended by those who believe that ``the best provider of information is a deregulated marketplace of ideas in which private media organizations compete for citizens' attention'' (378). Some legal scholars have recently applied these ideas to LLMs, which are undoubtedly new providers of information. As citizens ``increasingly turn to AI systems to easily access and aggregate information,'' citizens' right to receive speech justifies ``protecting such sources of information'' (Volokh et al. 2023: 654). Consequently, a law restricting ``AI output that is critical of the government, or that discusses abortion or gender identity or climate change (even if the restriction is framed in an ostensibly viewpoint-neutral way)'' would ``undermine users' ability to hear arguments that they might find persuasive and relevant to their political and moral decisionmaking'' (656).

However, Marciel notes that his argument leaves room for a different interpretation. The right to information might ground a positive duty for democratic governments to intervene more actively in the information landscape. Indeed, philosophers often doubt that a deregulated marketplace of ideas will really engender an informed citizenry. Elaborating on this idea, Habermas (1991: 228) writes that ``the formation of a public opinion in the strict sense is not effectively secured by the mere fact that anyone can freely utter his opinion and put out a newspaper.'' If so, democratic states could be obligated to create high-quality public media, assist non-profit organizations in creating private media, or even ``impose regulations on private media, such as the obligation to \emph{fairly} present different views on controversial issues'' (Marciel 2023: 378). On this interpretation, the idea that citizens possess a right to information could be used to justify a wide array of chatbot regulations: ones that prevent them from generating misinformation and conspiracy theories, ones that force them to answer user queries about controversial topics, ones that require the presentation of contrasting answers to the same question, etc.

Relatedly, legal scholars have argued that machine-generated speech should be protected even when no speaker's ability to speak is curtailed. In the U.S., Cass Sunstein (2023: 11) has written that ``any restriction on speech, even by an entity that lacks constitutional rights, must be adequately justified, if listeners or viewers claim that they want to hear or see the speech in question.'' According to his perspective, laws that prevent a chatbot from expressing a particular viewpoint would be unconstitutional. Sunstein grounds his argument in an analysis of \emph{Kleindienst v. Mandel}, 408 U.S. 753 (1972), a Supreme Court decision in which Justices acknowledged that the U.S. government's decision to deny a visa to Ernest Mandel, a Belgian Marxist invited to participate in a university colloquium, limited the First Amendment rights of those who wanted to \emph{listen} to Mandel. In Europe, Marco Bassini has underlined that Article 10 of the European Convention on Human Rights protects Europeans' freedom to ``receive and impart information and ideas without interference by public authority'' (Council of Europe 1950). Like Sunstein, he concludes that ``there is no apparent reason to exclude AI-generated content from constitutional coverage'' (Bassini 2025: 398).

\subsection{AI developers' corporate right to free speech}

The third potential holder of the right to free speech in the context of LLM-human interactions is AI developers. The philosophical issues at stake are not substantially different from those discussed in the first part of this chapter. If, as the U.S. Supreme Court has suggested in \emph{Netchoice}, social media platforms' decisions to ban certain kinds of speech constitute expressive conduct, then an AI company's decision to prevent its chatbot from generating certain kinds of speech arguably does as well. Like social media platforms, ``AI companies and their high-level employees may use autonomous agents and generative AI as a tool for guiding the creation of speech, consistently with their preferences and general ideologies'' (Volokh et al. 2023: 652). Undoubtedly, chatbots continuously generate speech with which their developers would likely disagree. However, they are still ``trained and refined to facilitate some forms of that content and inhibit others'' (653). What is more, chatbots are sometimes given a specific personality that reflects their developers' vision of what a meaningful interaction with an LLM should look like.~

Still, it is worth pausing to underscore the US-centricity of the preceding reasoning. Even if one argues that AI developers should be granted speech rights comparable to those of social media platforms, one can reasonably deny that implementing moderation guidelines counts as expressive conduct in the first place. For instance, a social media platform's moderation policy prohibiting spam or a chatbot guardrail that prevents a conversation from becoming too sexual might not be understood by users as conveying a particular message. This matters. Under the standard account established in \emph{Spence v. Washington}, 418 U.S. 405 (1974), expressive conduct must be intended to convey a message that an audience would recognize as such.\emph{~}

\subsection{How should we treat human-endorsed AI-generated speech?}

The preceding discussion focuses on cases in which a human interacts with an LLM-powered chatbot. In such cases, there is no audience involved (unless we envision the human user as an audience of one). In this section, I wish to consider philosophical questions raised by cases in which a human uses generative AI to produce speech, images, and videos, and then posts them elsewhere on the internet. When this happens, should this speech be treated the same as if the content had not been generated by artificial intelligence? Should it be treated as the user's speech and protected accordingly?~

Certainly, AI-generated content can convey meaning. Suppose, for instance, that someone creates a \textbf{deepfake} depicting Donald Trump kowtowing to Xi Jinping and posts it on \emph{TikTok}. This conveys the idea that, behind the scenes, the President of the United States is serving Chinese interests. Regardless of its truth value, this artifact clearly communicates a political message. The fact that a user produced it and posted it elsewhere also presumably indicates that they endorse or even identify with this message (unless, of course, they indicate otherwise). If so, then a content moderation policy that orders the removal of political deepfakes from \emph{TikTok} would frustrate the user's free speech interests. Of course, this does not mean that the deepfake should not be removed: as we have seen, many consider that platforms are within their rights to moderate what they consider to be bad speech (Messina 2024).

If platforms are within their right to moderate AI-generated content, how should they do so? Recently, philosophers have argued that, rather than developing new policies, platforms can apply existing ones to AI-generated text, images, or videos. Fisher, Howard, and Kira (2024: 132) explain:

\begin{quote}
\ldots fake audio of a politician is just as likely to disrupt the electoral process, whether it is made using generative AI or any other means; non-consensual intimate fakes are just as likely to cause psychological distress and reputational damage; and insinuating imagery is just as likely to denigrate a social group. Whatever content moderation policy a social media platform applies to these kinds of potentially harmful content when they are produced by humans, that policy should be applied across the board, in a technology-neutral way.
\end{quote}

This seems right, but it warrants commentary. Before the rise of generative AI, it was much harder for users to produce realistic images or videos that could effectively lead people to believe, for example, that Donald Trump \emph{really did} kowtow to Xi Jinping. As a result, the most relevant comparison is not between a highly realistic deepfake of Trump kowtowing and an equally realistic video of Trump kowtowing produced without AI. Instead, we should note that platforms now face a new challenge. Until recently, most users were limited to verbal accusations that Trump serves Chinese interests, but anyone can now falsely substantiate these claims by generating highly convincing AI imagery with ease. Before the generative AI revolution, platforms simply did not have policies addressing highly realistic fictional content, as such content was much rarer. It is therefore unsurprising that they are updating their policies to reflect changes in users' ability to mislead, for instance, by requiring that fake but realistic videos be labeled as such.~~

Interestingly, there is an ongoing philosophical discussion about such labels. Recently, Chomanski and Lauwaert (2025) have argued that labeling AI-generated political content should never be required by law, partly because this would entail a clear risk of governmental abuse. The specific risk is that some speech will be unfairly investigated by governments, depending on the viewpoint it expresses: ``the more effective such labels are at reducing credence in the labeled information, the more of a political incentive there is to scrutinize politically inconvenient messages more closely for signs of genAI use (and then insist on their labeling) than to scrutinize politically convenient messages'' (1005). Echoing Messina's analysis of governmental distrust justifications of free speech in this volume, the authors suggest that we should not expect governments, which are always composed of imperfect humans, to censor, sanction, or even label speech fairly.

This leaves open the question of whether platforms should voluntarily label synthetic content, as Fisher argues they should (2025). Specifically, Fisher rejects the idea that \emph{content generated by a user through an LLM} should be treated as \emph{the user's own speech}. In her view, users of LLMs ``lack sufficient control over the content produced to be deemed its (sole) author'' (280). Even if users control the LLM's output via prompts, ``many fine-grained decisions about the precise form of the output are still delegated to the model.'' On these grounds, she concludes that AI-generated content ``lacks the expressive value of human content,'' is not eligible for the same free speech protections, and should be clearly labeled to avoid potentially misleading audiences. Interestingly, Fisher falls short of arguing that these conclusions apply when a user ``adopts a piece of synthetic content as their own,'' for instance, by copy-pasting it into a social media post (281). When this happens, she suggests that the speech is attributable to said user and ``derives free speech value from their autonomy.''~

One risk of this caveat is that it may undermine Fisher's broader case for labeling. If copying and pasting text into a social media post counts as adopting speech, why would posting a deepfake on social media not count as adopting it? What, precisely, counts as adopting someone's speech as my own? In general, discussions about AI-generated content raise questions at the intersection of several philosophical disciplines, including ethics, political philosophy, philosophy of mind, and philosophy of language. Such discussions also invite us to revisit the nature and meaning of authorship, an issue I prefer, for space reasons, to leave to other philosophers.~

\textbf{Conclusion}

Artificial intelligence is a deeply ambiguous term. In recent decades, it has been used to refer to a wide range of technologies capable of replacing human cognition and action. As a result, discussing the challenges AI poses to freedom of expression is difficult; these challenges vary depending on the specific AI technology in question. In the preceding discussion, I examined the free speech issues raised by two such technologies: recommendation algorithms that structure social media feeds and LLM-powered chatbots. As we have seen, the free speech questions raised by recommendation algorithms are a natural extension of those raised by content moderation and, more generally, by the private governance of speech. When discussing these questions, my central claim has been that the value of speech largely derives from its \emph{visibility}, and that recommendation algorithms are among the main drivers of visibility and invisibility in the contemporary digital public sphere.

By contrast, chatbots raise distinct philosophical questions, primarily because users interact with them without an audience present. As we have seen, a key question raised by the possibility that states will regulate this technology concerns the nature and scope of users' right to seek and receive information.~On this issue, those who argue that users' right to information should be protected from state interference offer arguments that resonate with the thinker-based approach to freedom of expression. According to this view, the value of a speech right derives from its capacity to enable persons to become autonomous agents by developing their emotional and rational capacities. Through communication, human beings participate in joint endeavors, develop moral agency, forge meaningful relationships, and come to understand themselves. Whether LLMs can enable thinkers to acquire all those capacities remains an open question, but categorically ruling out that they can help them develop at least some of them would amount to hasty reasoning. Consequently, philosophers interested in expression and communication have already begun turning their attention to AI, and there are few reasons to believe they should refrain from doing so.~

\textbf{Questions for discussion}

\begin{enumerate}
\def\labelenumi{\arabic{enumi}.}
\item
  On social media platforms, recommendation algorithms can make user-generated speech highly visible or nearly invisible. In democratic societies, do people have a right against attempts to make their speech invisible? If so, should this right constrain the behavior of social media platforms?
\item
  Should we -- derivatively or non-derivatively -- attribute speech rights to AI agents? Should we attribute rights to them at all? What capacities would they need to possess to be rights holders?
\item
  Do people possess a right to information? If so, what is the scope of the right, and what does it entail for the government regulation of large language models?
\item
  When a person generates an image or video using artificial intelligence and posts it online, should it be treated as that person\textquotesingle s own speech? Does it matter how much control they had over what the model produced?
\end{enumerate}

\section*{References}
\begin{hangparas}{2em}{1}
\setlength{\parskip}{6pt}

Austin, J. L. (1975)~\emph{How to Do Things with Words}. 2nd edn. Oxford: Clarendon Press.

Basl, J. and Bowen, J. (2020) "AI as a Moral Right-Holder," in M. D. Dubber, F. Pasquale, and S. Das (eds.),~\emph{The Oxford Handbook of Ethics of AI}. Oxford University Press: 289--306.

Bassini, M. (2025) "Speech without a Speaker: Constitutional Coverage for Generative AI Output?,"~\emph{European Constitutional Law Review}~21 (3): 375--411.

Benjamin, S. (2013) "Algorithms and Speech,"~\emph{University of Pennsylvania Law Review}~161 (6): 1445--1493.

Bonotti, M. and Seglow, J. (2022) "Freedom of Speech: A Relational Defence,"~\emph{Philosophy \& Social Criticism}~48 (4): 515--529.

Bousquet, C. (2025) "Limiting the Right to Moderate: Political Equality, Social Media, and Viewpoint-Based Moderation,"~\emph{Ethics and Information Technology}~27 (4): 62.

Brown, É. (2023) "Free Speech and the Legal Prohibition of Fake News,"~\emph{Social Theory and Practice}~49 (1): 29--55.

------ (2025) "Recommended Selves: Authenticity and Algorithmic Filtering,"~\emph{Journal of the American Philosophical Association}~(FirstView): 1--20.

------ (2026) "Democracy Needs Reach: Political Equality, Online Speech and Algorithmic Recommendation,"~\emph{Ethical Theory and Moral Practice}~29 (2): 433--449.

Butlin, P. and Viebahn, E. (2025) "AI Assertion,"~\emph{Ergo}~12 (37): 968--988.

Calvet-Bademunt, J. and Mchangama, J. (2024)~\emph{Freedom of Expression in Generative AI: A Snapshot of Content Policies}. The Future of Free Speech. Available at:~\url{https://futurefreespeech.org/wp-content/uploads/2023/12/FFS_AI-Policies_Formatting.pdf}~(accessed 2 July 2026).

Chomanski, B. and Lauwaert, L. (2025) "Automated Propaganda: Labeling AI-Generated Political Content Should Not Be Required by Law,"~\emph{Journal of Applied Philosophy}~42 (3): 994--1015.

Council of Europe (1950)~\emph{European Convention for the Protection of Human Rights and Fundamental Freedoms}. Available at:~\url{https://www.echr.coe.int/documents/convention_eng.pdf}~(accessed 30 June 2026).

Dworkin, R. (2010) "Foreword," in I. Hare and J. Weinstein (eds.),~\emph{Extreme Speech and Democracy}. Oxford University Press: v--ix.

Fisher, S. A. (2025) "Something AI Should Tell You: The Case for Labelling Synthetic Content,"~\emph{Journal of Applied Philosophy}~42 (1): 272--286.

Fisher, S. A., Howard, J. W., and Kira, B. (2024) "Moderating Synthetic Content: The Challenge of Generative AI,"~\emph{Philosophy \& Technology}~37 (4): 133.

FitzGerald, K. M. (2025) "AI Chatbots Are Encouraging Conspiracy Theories -- New Research,"~\emph{The Conversation}, 24 November.

Garvey, J. (2022) "Let\textquotesingle s Get Real: Weak Artificial Intelligence Has Free Speech Rights,"~\emph{Fordham Law Review}~91 (3): 953.

Gillespie, T. (2022) "Do Not Recommend? Reduction as a Form of Content Moderation,"~\emph{Social Media + Society}~8 (3).

Goldman, E. (2024) "Bonkers Opinion Repeals Section 230 in the Third Circuit -- Anderson v. TikTok,"~\emph{Technology and Marketing Law Blog}, 24 August. Available at:~\url{https://blog.ericgoldman.org/archives/2024/08/bonkers-opinion-repeals-section-230-in-the-third-circuit-anderson-v-tiktok.htm}~(accessed 30 June 2026).

Grafanaki, S. (2018) "Platforms, the First Amendment and Online Speech: Regulating the Filters,"~\emph{Pace Law Review}~39 (1): 111.

Habermas, J. (1991)~\emph{The Structural Transformation of the Public Sphere: An Inquiry into a Category of Bourgeois Society}. Cambridge, MA: MIT Press.

Haidt, J. (2001) "The Emotional Dog and Its Rational Tail: A Social Intuitionist Approach to Moral Judgment,"~\emph{Psychological Review}~108 (4): 814--834.

Hinsliff, G. (2025) "When a Chatbot\textquotesingle s Advice Is a Matter of Life or Death, How Can We Leave AI to the Free Market Wild West?,"~\emph{The Guardian}, 9 December. Available at:~\url{https://www.theguardian.com/commentisfree/2025/dec/09/would-you-entrust-a-childs-life-to-a-chatbot-thats-what-happens-every-day-that-we-fail-to-regulate-ai}~(accessed 30 June 2026).

Jannach, D., Manzoor, A., Cai, W., and Chen, L. (2022) "A Survey on Conversational Recommender Systems,"~\emph{ACM Computing Surveys}~54 (5): 1--36.

Kirby, N. (2026) "Divide and Profit: Affective Polarisation, Social Media Regulation and Free Speech." Unpublished manuscript.

Kramer, M. H. (2021)~\emph{Freedom of Expression as Self-Restraint}. Oxford: Oxford University Press.

Lazar, S. (2025) "Governing the Algorithmic City,"~\emph{Philosophy \& Public Affairs}~53 (2): 102--168.

List, C. (2021) "Group Agency and Artificial Intelligence,"~\emph{Philosophy \& Technology}~34 (4): 1213--1242.

Liu, N., Hu, X. E., Savas, Y., et al. (2025) "Short-Term Exposure to Filter-Bubble Recommendation Systems Has Limited Polarization Effects: Naturalistic Experiments on YouTube,"~\emph{Proceedings of the National Academy of Sciences}~122 (8): e2318127122.

Marciel, R. (2023) "On Citizens\textquotesingle{} Right to Information: Justification and Analysis of the Democratic Right to Be Well Informed,"~\emph{Journal of Political Philosophy}~31 (3): 358--384.

Massaro, T. and Norton, H. (2016) "Siri-ously? Free Speech Rights and Artificial Intelligence,"~\emph{Northwestern University Law Review}~110 (5): 1169--1194.

Messina, J. P. (2020) "Freedom of Expression and the Liberalism of Fear: A Defense of the Darker Mill,"~\emph{Philosopher\textquotesingle s Imprint}~20 (34): 1--17.

------ (2024)~\emph{Private Censorship}. Oxford: Oxford University Press.

Miller, E. (2021) "Amplified Speech,"~\emph{Cardozo Law Review}~43 (1): 1--70.

Narayanan, A. (2023) "Understanding Social Media Recommendation Algorithms,"~\emph{Knight First Amendment Institute}, 9 March. Available at:~\url{http://knightcolumbia.org/content/understanding-social-media-recommendation-algorithms}(accessed 30 June 2026).

Ovadya, A. and Thorburn, L. (2023) "Bridging Systems: Open Problems for Countering Destructive Divisiveness across Ranking, Recommenders, and Governance,"~\emph{Knight First Amendment Institute}, 26 October. Available at:~\url{https://knightcolumbia.org/content/bridging-systems}~(accessed 30 June 2026).

Parekh, B. (2012) "Is There a Case for Banning Hate Speech?," in M. Herz and P. Molnar (eds.),~\emph{The Content and Context of Hate Speech: Rethinking Regulation and Responses}. Cambridge: Cambridge University Press: 37--56.

Schauer, F. (2002) "First Amendment Opportunism,"~in L. C. Boilinger and G. R. Stone, \emph{Eternally Vigilant: Free Speech in the Modern Era}, Chicago: University of Chicago Press: 174--197.

Schwitzgebel, E. (2025) "AI and Consciousness." arXiv preprint, arXiv:2510.09858.

Shiffrin, S. V. (2014)~\emph{Speech Matters: On Lying, Morality, and the Law}. Princeton, NJ: Princeton University Press.

Solum, L. (1992) "Legal Personhood for Artificial Intelligences,"~\emph{North Carolina Law Review}~70 (4): 1231.

Sunstein, C. R. (2023) "Artificial Intelligence and the First Amendment,"~\emph{SSRN Electronic Journal}.

Volokh, E. and Falk, D. M. (2012) "First Amendment Protection for Search Engine Search Results,"~\emph{UCLA School of Law Research Paper}~No. 12-22.

Volokh, E., Lemley, M. A., and Henderson, P. (2023) "Freedom of Speech and AI Output,"~\emph{Journal of Free Speech Law}~3: 651--660.

West, A., Novelli, C., Taddeo, M., and Floridi, L. (2025) "Recommender Systems as Commercial Speech: A Framing for US Legislation,"~\emph{Ethics and Information Technology}~27 (4): 52.

Wu, T. (2013) "Machine Speech,"~\emph{University of Pennsylvania Law Review}~161 (6): 1495.
\end{hangparas}

\end{document}